\documentstyle[12pt] {article}
\newcommand{\alpi}{\it O(\alpha/\pi)}
\def\beq{\begin{equation}}
\def\eeq{\end{equation}}
\def\amu{a_\mu}
\def\maxmodamu{max~|a_\mu^{SUSY}|}
\def\tanbeta{{\rm tan}\beta}
\def\gmin2{(g-2)_\mu}
\def\ecoup{{e \over {{\sqrt2}\sin{\theta_W}}}}
\def\winomass{m_{\tilde W_a}}
\def\zinomass{m_{\tilde Z_{(k)}}}
\def\denom{{\left[4{\nu}^2_{1,2}+{({\mu} \pm {\tilde m}_2)}^2 \right]}
^{-\frac{1}{2}}}

\def\BR{B_k^R}
\def\BL{B_k^L}
\def\Gff{G_1(x_{1k})}
\def\Gfs{G_1(x_{2k})}
\def\Gsf{G_2(x_{1k})}
\def\Gss{G_2(x_{2k})}
\catcode`\@=11 % This allows one to modify PLAIN macros.

\def\lsim{\mathrel{\mathpalette\@versim<}}
\def\gsim{\mathrel{\mathpalette\@versim>}}
\def\@versim#1#2{\vcenter{\offinterlineskip
    \ialign{$\m@th#1\hfil##\hfil$\crcr#2\crcr\sim\crcr } }}

\def\PL{Phys. Lett.}
\def\PRL{Phys. Rev. Lett.}
\def\PHYREP{Phys. Rep.}
\def\NP{Nucl. Phys.}
\def\PR{Phys. Rev.}
\def\ZPHY{{Z. Phys C} }

\begin{document}
\begin{flushright}
{NSF-ITP-95-64}\\
{NUB-TH-3125/95}\\
%\date{\today}
\end{flushright}
%-----------------------------------
%\documentstyle[preprint,aps]{revtex}
\begin{center}
{\Large\bf Probing Supergravity Grand Unification \\in the Brookhaven
g-2 Experiment\\}
\vglue 0.5cm
{\large U. Chattopadhyay$^{(a)}$ and Pran Nath$^{(a)}\footnote{Permanent
address}
^{,(b)}$}
\vglue 0.2cm
{\em $^{(a)}$Department of Physics, Northeastern University, Boston,
MA 02115, USA\\}
{\em $^{(b)}$Institute for Theoretical Physics, University of California}\\
{\em Santa Barbara, CA 93106-4030, USA\\}
\end{center}

\begin{abstract}
A quantitative analysis of $\amu\equiv{1\over 2}(g-2)_\mu$ within
the framework of Supergravity Grand Unification and  radiative
breaking of the electro-weak symmetry  is given.
It is found that $a_{\mu}^{SUSY}$ is dominated by the chiral interference
	term from the light chargino exchange, and that this term carries a
signature which correlates strongly with the sign of $\mu$. Thus as a rule
  $a_{\mu}^{SUSY}>0$ for $\mu>0$ and
 $a_{\mu}^{SUSY}<0$  for $\mu<0$ with very few exceptions when tan$\beta\sim
1$.
At the quantitative level it is
shown that if the E821 BNL experiment can reach the expected
sensitivity of $4\times 10^{-10}$ and there is a reduction in
the hadronic error by a factor of four or more, then the
experiment  will explore a majority of the parameter space
in $ m_0-m_{\tilde g}$ plane in the region $m_0\lsim 400$ GeV,
$m_{\tilde g}\lsim 700$ GeV for $\tanbeta \gsim 10$ assuming the
experiment will not discard the Standard Model result within its
$2\sigma$ uncertainty limit. For smaller
$\tanbeta$, the SUSY reach of E821 will still be considerable.
 Further, if no effect within $2 \sigma$ limit of the Standard
Model value is seen, then large $\tanbeta$ scenarios
 will be severely constrained within
the current naturalness
criterion, ie.,  $m_0, m_{\tilde g}\lsim 1$ TeV.
\end{abstract}
%--------------------

\section{ Introduction}

  The high level of experimental accuracy of the measured value of the
anomalous magnetic moment of the muon~\cite{bailey}
has provided  verification
to several orders in the perturbation expansion
of quantum electrodynamics(QED)~\cite{qedrefs} as well as put constraints
on physics beyond the Standard Model~(SM)~\cite{km}. Further
the E821 experiment currently underway at Brookhaven is expected to improve
the accuracy over the previous measurement by a factor of 20~\cite{hughes}.
Simultaneously,
it is expected that improved analyses of existing data on
$(e^+e^-\rightarrow hadrons)$ as well as new data from
ongoing experiments in VEPP-2M~\cite{veppref} together
 with future experiments in DA$\Phi$NE~\cite{daphineref},
BEPC~\cite{bepcref} etc. will reduce the
uncertainty  in the hadronic contributions to a significant level so as to
allow for a test for the first time of the elctro-weak corrections in the
Standard Model.
	It was pointed out in Refs~\cite{yuan,kosower}
(~see also Refs~\cite{fayet} for a discussion of previous work and
Ref~\cite{lopez} for a discussion of recent work~)
	that supersymmetric corrections to
$\gmin2$ are in general the same size as the electro-weak corrections in
supergravity grand unification~\cite{chams}. Thus improved experiments
designed to
test the Standard Model electro-weak corrections can also provide a probe
 of supersymmetric contributions.

\begin{center} \begin{tabular}{|l|c|c|}
\multicolumn{3}{c}{Table~1: Contributions to $ a_\mu \times
10^{10} $ } \\
%\hline
\hline
Nature of Contribution &CASE A  &CASE B \\
\hline
Q.E.D. to $\alpi^5$  &11658470.8(0.5)  &11658470.8(0.5) \\
Kinoshita {\it et al.}~\cite{czar,qedrefs} &  &  \\
\hline
Hadronic vac. polarization  &705.2(7.56) &702.35(15.26) \\
to $\alpi^2$ & Martinovic \& Dubnicka~\cite{martinov}  &
Eidelman \& Jegerlehner~\cite{eidel,worstell} \\
% &  & \\
Hadronic vac. polarization  &$-9.0(0.5)$ &$-9.0(0.5)$\\
to $\alpi^3$ &  & \\
%Kinoshita {\it et al.} &  & \\
Kinoshita \& Marciano~\cite{km}  &  & \\
% &  & \\
Light by light hadronic  &0.8(0.9) &0.8(0.9) \\
amplitude &  & \\
Bijnens {\it et al.}~\cite{bijn} &  & \\
% &  & \\
Total hadronic  &697.0(7.6) &694.2(15.3) \\
\hline
Electro-weak one-loop  &19.5   &19.5 \\
Fujikawa {\it et al.}~\cite{fuji}  &       &     \\
% &  & \\
Electro-weak 2-fermion loops  &$-2.3(0.3)$ &$-2.3(0.3)$ \\
Czarnecki {\it et al.}~\cite{czar}  &  & \\
%  &  & \\
Electro-weak 2-boson loops  &$(-2.0+0.045 \times R_b)$  &$(-2.0+0.045
\times R_b)$ \\
Czarnecki {\it et al.}~\cite{czar} &  & \\
% &  & \\
Total electro-weak   &$(15.2(0.3)+0.045 \times R_b)$
 &$(15.2(0.3)+0.045 \times R_b)$ \\
upto 2-loops  &  & \\
\hline
\hline
Total (with ${\rm R_b=0}$)  &11659183(7.6) &11659180(15.3) \\
\hline
%\hline
\end{tabular}
\end{center}

\noindent
The analyses of Refs~\cite{yuan,kosower}, however,
were done without using the constraints
of radiative breaking of the electro-weak symmetry~\cite{inoue}
and without the benefit
	of the recent high precision LEP data~\cite{davier,erler}
to constrain
the coupling constants~\cite{unification}.
The purpose of the present work is to
include these features in
	the  analysis. Additionally, we investigate the effects on the results
due to $b\rightarrow s\gamma$ constraint~\cite{cleo}, and dark matter
 constraint.
We  shall find that the  expected accuracy of
 $\amu\equiv{1\over 2}(g-2)_\mu$,
in E821 BNL experiment would either allow for supergravity grand unification
	effects to be visible in the BNL experiment, or if no effect
	beyond the $2\sigma$ level is seen, then there will be a
significant constraint on the model.

  The most recent experimental value of $\amu $
is averaged to be  $ \amu^{exp}=1165923(8.4)\times 10^{-9} $.
The quantity within the parenthesis refers to the uncertainty in the last
digit.
The Standard Model results consist of several parts:

\beq
a_{\mu}^{SM}=a_{\mu}^{qed}+a_{\mu}^{had} + a_{\mu}^{E-W}
\eeq
Here $a_{\mu}^{qed}$ \cite{qedrefs} is computed to
$O(\alpha^5)$ QED corrections.
$ a_{\mu}^{had}$ consists of $O(\alpha^2)$~\cite{martinov,eidel} and
$O(\alpha^3)$~\cite{km} hadronic
vacuum polarizations, and also light-by-light hadronic
contributions~\cite{bijn}.
We exhibit two different evaluations in Table~1. Case~A  uses the analysis
of Martinovic and Dubnicka~\cite{martinov} who use a
detailed structure of pion and kaon
form factors in their fits to get the $O(\alpha^2)$ corrections. Case~B uses
the $O(\alpha^2)$ analysis of Eidelman and Jegerlehner~\cite{eidel,worstell}.
We note that there is
almost a factor of two difference between the hadronic errors of case~A and
case~B. For the electro-weak corrections we have used the recent analyses
of Czarnecki et al.~\cite{czar}, which include one-loop electro-weak
corrections
of the Standard Model~\cite{fuji},
2-loop corrections with fermion loops~\cite{czar,peris}, and partial
2-loop corrections with boson loops~\cite{kkss}. The
remaining unknown 2-loop
corrections with boson loops are denoted by $R_b$ following the notation of
Ref.~\cite{czar}. Pending the  full 2-loop
bosonic contributions in the E-W corrections, the total $a_\mu^{SM}$
for case~A
is 11659180(7.6)$\times 10^{-10}$ while for case~B one has 11659200(15.3)
 $\times 10^{-10}$. We see that the overall error, which is dominated by the
hadronic corrections, is about half for case~A relative to that for case~B.

 The new high precision E821 Brookhaven experiment~\cite{hughes}
 has an anticipated
design sensitivity of $4\times 10^{-10}$.  This is about 20 times
more accurate than that of the CERN measurement conducted earlier.
  However, as mentioned already one needs in addition an improvement
 in the computation of
the hadronic contribution $\amu^{had}$, where uncertainties primarily
arise due to hadronic
vacuum polarization effects. This problem is expected to be overcome
soon through further accurate measurements of $\sigma_{tot}(e^+e^-
\rightarrow hadrons)$ for the low energy domain.

\section{ The Minimal Supergravity Model}

  The framework of our analysis is N=1 Supergravity grand unified
theory~\cite{chams} where the
Supergravity interactions spontaneously break supersymmetry at the Planck scale
($M_{Pl}=2.4\times 10^{18}$ GeV)  via a hidden sector~\cite{chams}.
Further we assume that the GUT group G breaks at scale $Q=M_G$ to the
 Standard Model gauge group :
$G \longrightarrow SU(3)_C\times SU(2)_L\times U(1)_Y$.
  At low energy the symmetry breaking effective potential below the GUT scale
$M_G$ is given by

\beq
 V_{SB}=m^2_0z_iz_i^\dagger + (A_0W^{(3)}+B_0W^{(2)}+h.c.)
\label{pot}
\eeq
where $W^{(2)}, W^{(3)}$ are the quadratic and the cubic parts of the
effective superpotential.
  Here $W^{(2)}=\mu_0H_1H_2$, with $H_1$, $H_2$ being two Higgs
doublets, and $W^{(3)}$ contains terms cubic in fields and involves
the interactions of
quarks, leptons and Higgs with strength determined by Yukawa couplings.
In addition the low energy
theory has a universal gaugino mass term $-m_{\frac{1}{2}}\bar\lambda^\alpha
 \lambda^\alpha$.
  The minimal Supergravity model below the GUT scale depends on the
following set of parameters.
\beq
m_0, m_{\frac{1}{2}}, A_0, B_0; \mu_0, \alpha_G, M_G
\eeq
where $M_G$ and $\alpha_G$ are the GUT mass and the coupling constant
respectively.
  Among these parameters $\alpha_G$ and $M_G$ are determined with the high
precision LEP results by using two-loop renormalization group equations
$\alpha_i(M_Z), i=1,2,3$ up to $M_G$~\cite{einhorn}.
Renormalization group analysis is used to break the electro-weak
symmetry~\cite{inoue} and radiative breaking allows one to
determine $\mu_0$ by fixing
$M_Z$ and to find $B_0$ in terms of
$tan\beta={{\langle H_2 \rangle}\over {\langle H_1 \rangle}}$.
  Thus the model is completely parametrized by just four quantities~\cite
{scaling}
\beq
  m_0, m_{\tilde g}, A_t, tan\beta
\eeq
  Here the universal gaugino mass
$m_{\frac{1}{2}}$ is replaced by the gluino mass $m_{\tilde g}$
and $A_0$ by $A_t$, which is the t-quark A parameter at
the electro-weak scale $M_{EW}$. In addition to these four parameters
and the top quark mass $m_t$, one has to specify the sign of $\mu$, since
the radiative breaking  equations determine only $\mu^2$.
   There are 32 new particles in this model (12 squarks, 9 sleptons, 2
charginos, 4 neutralinos, 1 gluino, 2 CP even neutral Higgs, 1 CP odd
neutral Higgs and 1 charged Higgs). The masses of these 32 new particles
and all their interactions can be determined by the four parameters
mentioned above. Thus the theory makes many new
predictions~\cite{scaling}, and has led to considerable activity~\cite
{ross,bagger}
 to explore the implications of supergravity grand unification and its
signals~\cite{baer}. The allowed parameter space of the model is
further constrained
by the following i) charge and colour conservation~\cite{frere},
ii) absence of tachyonic particles,
iii) a lower bound on SUSY particle masses as indicated by CDF, D0, and LEP
data,iv) an upper limit on SUSY masses from the naturalness criterion which is
taken as  $m_0, m_{\tilde g}< 1$~TeV.Our analysis automatically takes
into account important Landau pole effects that arise due the top being
heavy and thus in proximity to the Landau pole that lies in the top
Yukawa coupling~\cite{landau}. Additionally, we  consider the
constraint on
 $b \rightarrow s\gamma$ decay rate from the recent CLEO data, and
 the neutralino relic density constraint.

\section{Analysis of $(g-2)_\mu^{SUSY}$ and Results}

  We use Ref.~\cite{yuan} to compute SUSY contributions to $\gmin2$.
These contributions arise from Figs.~(1a) and ~(1b). In Fig.~(1a) one
exchanges 2 charginos
${\tilde W}_a$, a=1,2 with masses $\winomass$ which are charged spin
${\frac {1}{2}}$ Dirac fields and a sneutrino state.
In Fig.~(1b) one exchanges 4 neutralinos
${\tilde Z}_{(k)}$, k=1,2,3,4 with masses $\zinomass$
which are
spin ${\frac {1}{2}}$ Majorana fields (our labelling of particles satisfy
${\tilde m}_i < {\tilde m}_j$ for $i<j$), and two scalar smuon mass
eigenstates.
 The mass
spectra of the charginos, neutralinos, smuons and of the sneutrino
is given in the Appendix for convenience. The one-loop supersymmetric
contribution
to $(g-2)_\mu$  then is given by
\beq
\amu^{SUSY}={\frac{1}{2}}\gmin2^{SUSY}=
{\frac{1}{2}}{\left(g^{\tilde W}+g^{\tilde Z}\right)}
\eeq
For the chargino-sneutrino part referring to~\cite{yuan} we find
(with summation over repeated indices implied)
\beq
 g^{\tilde W}={{m^2_\mu} \over {24{\pi}^2}} {{ {A_R^{(a)}}^2} \over
{\winomass^2}}F_1(x_a)+{{m_\mu} \over{4{\pi}^2}} {{A_R^{(a)}
A_L^{(a)}} \over {\winomass}} F_2(x_a)
\label{winoeqn}
\eeq

\beq
x_a=\left({{m_{\tilde \nu}} \over {m_{\tilde W_a}}}\right)^2; \quad a=1,2
\label{winosecond}
\eeq
The mass of the  sneutrino required in Eq.~(\ref{winosecond}) can be
determined from Eq.~(\ref{sneutrinoeqn}).
The first term in Eq.~(\ref{winoeqn}) contains terms diagonal in chirality
whereas the second term has R-L interference terms arising from Yukawa
interactions. The functions
$F_1(x)$ and $F_2(x)$ are defined by
\begin{eqnarray}
 F_1(x) & =& (1-5x-2x^2)(1-x)^{-3} -6x^2(1-x)^{-4}ln(x)  \nonumber  \\
 F_2(x) & =& (1-3x)(1-x)^{-2}-2x^2(1-x)^{-3}ln(x)
\end{eqnarray}
$A_R^{(a)}$ and $A_L^{(a)}$ of Eq.~(\ref{winoeqn}) are given as
\beq
A_R^{(1)}=-{\ecoup}\cos {\gamma_1}; \quad A_L^{(1)}={(-1)^\theta}
{\ecoup}{{{m_\mu}\cos{\gamma_2}} \over {{\sqrt2}M_W \cos {\beta}}}
\label{ALAR}
\eeq

\beq
A_R^{(2)}=-{\ecoup}\sin {\gamma_1}; \quad A_L^{(2)}=-{\ecoup}
{{{m_\mu}\sin{\gamma_2}} \over
{{\sqrt2}M_W \cos {\beta}}}
\eeq
Here the angles $\gamma_1$ and $\gamma_2$ are found by using
${\gamma_{1,2}}={\tilde {\beta}}_2 \mp {\tilde {\beta}}_1$
where
\begin{eqnarray}
{\sin2{{\tilde {\beta}}_{1,2}}} & = &{{\left(\mu \pm {\tilde m}_2 \right)}}
{\denom} \nonumber \\
{\cos2{{\tilde {\beta}}_{1,2}}} & = & 2{\nu}_{1,2}{\denom}
\end{eqnarray}
  where ${-\pi <2{\tilde {\beta}}_{1,2} \leq {\pi}}$.

  The neutralino-smuon loop correction results in

\begin{eqnarray}
g^{\tilde Z} & &=-{{m^2_\mu} \over {12{\pi}^2}}{1 \over {\zinomass^2}}
{\left[{\lbrace s^2 {(\BR)}^2 +c^2 {(\BL)}^2\rbrace}\Gff +
{\lbrace c^2 {(\BR)}^2 +s^2 {(\BL)}^2\rbrace}\Gfs\right]}\nonumber\\
& &
\hspace{4cm} +{{m_\mu} \over {4\pi^2}}{sc\over {\zinomass}}{\BR \BL}
{\left[\Gsf - \Gss\right]}\nonumber\\
& &
-{{m_\mu} \over {4\pi^2}}{{C_k}\over \zinomass}
{\left[{\lbrace c^2\BL -{(-1)^{\theta_k}}s^2\BR\rbrace}\Gsf
+ {\lbrace s^2\BL -{(-1)^{\theta_k}}c^2\BR \rbrace}\Gss\right]}
\label{zinoeqn}
\end{eqnarray}
Here $s=\sin \delta$, $c=\cos \delta$, and
\beq
x_{rk}={\left({m_{\tilde {\mu_r}} \over {\zinomass}}\right)}^2; \quad r=1,2;
 \quad k=1,2,3,4
\label{smuonfirst}
\eeq
The functions $G_1(x)$ and $G_2(x)$ are given by
\begin{eqnarray}
G_1(x) & = & {\frac{1}{2}}(2+5x-x^2){(1-x)}^{-3}+3x{(1-x)}^{-4}ln(x)
\nonumber \\
G_2(x) & = & (1+x){(1-x)}^{-2} + 2x{(1-x)}^{-3}ln(x)
\end{eqnarray}
The Yukawa coefficients $C_k$ are found from
\beq
C_k=e{{m_\mu} \over {2M_W{\cos\beta}{\sin\theta_W}}}
{\left[{\cos\beta}O_{3k}+{\sin\beta}O_{4k} \right]}
\eeq
 and $\BR$ and $\BL$ are found by using

\begin{eqnarray}
{\BR} & = & e\left[-O_{1k} +\cot2{\theta_W}O_{2k} \right] \nonumber \\
{\BL} & = & e\left[-O_{1k} -\tan{\theta_W}O_{2k} \right]{(-1)}^{\theta_k}
\label{brbleqn}
\end{eqnarray}
The angle $\delta$, $\theta_k$ and the quantities
 $O_{ij}$ are defined in the Appendix.

  Among the two sources of one-loop contributions to
$\amu^{SUSY}$
the chargino-sneutrino loop contributes more than
the neutralino-smuon loop. This occurs mainly due to the
smallness of the mixing angle $\delta$ (see Eq.~(\ref{zinoeqn})) which
itself arises from the smallness of the muon to the sparticle mass ratio
 (see Eq.~(\ref{deleqn})).
Partial cancellations on the right hand side of Eq.~(\ref{zinoeqn})
are also responsible
for a reduction of the  neutralino-smuon contribution.
For the chargino-sneutrino contribution (see Eq.~(\ref{winoeqn}))
one has comparable magnitudes for chirality diagonal
and non-diagonal terms for small $\tanbeta (\sim 1)$.  The
non-diagonal terms become more important as $\tanbeta$ starts to
deviate significantly from unity.
Indeed, a large value of $\tanbeta$ results in a large contribution of
the chirality non-diagonal term in the chargino-sneutrino part and
hence to $|\amu^{SUSY}|$  due to the enhancement  arising from
$1 \over{{\rm cos}\beta}$ $\sim \tanbeta$ in the Yukawa coupling.

 There exists a very strong  correlation between the sign of
$\amu^{SUSY}$ and the sign of $\mu$ which we now explain.
  While the chiral interference chargino part of $\amu^{SUSY}$ dominates
over the other terms, it is the lighter chargino mass which
 contributes  most dominantly. In fact this part depends on
$A_L^{(1)}A_R^{(1)}$ which from Eq.~(\ref{ALAR}) can be seen to have
a front factor of
 $(-1)^\theta$, where $\theta=0(1)$ for
$\lambda_1>0(<0)$ where
$\lambda_1$ is the smaller eigenvalue of the chargino mass matrix (see
 Eq.~(\ref{lambdaequation})). For
$\mu>0$, one finds $\lambda_1<0$ invariably and for $\mu<0$ one has
$\lambda_1>0$ for almost all the regions of parameter space of interest.
This can be seen by writing $\lambda_{1,2}$ in the following form
\begin{equation}
{\lambda_{1,2}}={\frac{1}{2}}\left(
{\left[2M^2_W+{({\mu}-{\tilde m}_2)}^2-2M^2_Wsin2\beta
\right]}^{\frac{1}{2}} \mp
{\left[2M^2_W+{({\mu}+{\tilde m}_2)}^2+2M^2_Wsin2\beta
\right]}^{\frac{1}{2}}
\right)
\label{winonew}
\end{equation}
and noting  that the terms containing
${\rm \sin}2\beta$ are only appreciable when $\tanbeta$ is small
$(\sim 1)$.
 As a result, due to this unique dominance of the chiral interference
term involving the lighter chargino mass, one finds as a rule $\amu^{SUSY}>0$
for $\mu>0$ and $\amu^{SUSY}<0$ for $\mu<0$ with some very few
exceptions for the latter case when $\tanbeta \sim 1$ resulting in a very
small $|\amu^{SUSY}|$.

 The generic dependence of $|\amu^{SUSY}|$ on the remaining parameters
$m_0$, $m_{\tilde g}$, and $A_t$ is as follows: Regarding $m_0$,
the dependence results primarily from
the chargino-sneutrino sector because the mass spectra depend on $m_0$.
This results in a decreasing  $|\amu^{SUSY}|$  with increase in $m_0$.
  Among the other two basic parameters a large gluino mass $m_{\tilde g}$
in general again leads to a smaller $|\amu^{SUSY}|$, due to
the resulting larger sparticle masses entering in the loop.
The $A_t$ dependence enters implicitly via the SUSY mass spectra, and
also explicitly in the neutralino-smuon exchange diagrams.
  Figs.~(2a) and (2b) show the upper limit of $|\amu^{SUSY}|$ in the
minimal Supergravity model by mapping
 the entire parameter space for
$m_0, m_{\tilde g}\leq 1$~TeV, $\tanbeta \leq 30$ and both signs of $\mu$.
 One finds that $\maxmodamu$ increases with increasing
 $\tanbeta$ for fixed $m_{\tilde g}$.
  As discussed above this happens due to the essentially linear
dependence of the chirality non-diagonal term on $\tanbeta$.
Furthermore, for large $\tanbeta$ one finds similar magnitudes of
$\maxmodamu$ for $\mu>0$ and $\mu<0$. This can be understood
by noting that for a large $\tanbeta$ the lighter chargino  masses
for $\mu>0$ and $\mu<0$ cases have almost similar magnitudes
(see ~Eq.~(\ref {winonew})).
This, along
with the explanation of the dependence of the sign of $\amu^{SUSY}$ on the
sign of $\mu$ accounts for similar $|\amu^{SUSY}|$ values for both
$\mu<0$ and $\mu>0$ when $\tanbeta$ is large. This similarity between the
$\mu>0$ and $\mu<0$ cases  is less apparent for small
$\tanbeta$ (ie. $\tanbeta\lsim 2$) and small $m_{\tilde g}$ values.

Next we include in the analysis $b \rightarrow s\gamma$ constraint and the dark
matter constraint which have been shown in recent
work~\cite{bertolini,wu} to generate
strong constraints on the parameter space. For the $b \rightarrow s\gamma$
decay the CLEO Collaboration~\cite{cleo} gives a value of

\beq
{\rm BR}(b \rightarrow s\gamma)=(2.32\pm 0.5\pm 0.29\pm 0.32)\times 10^{-4}
\label{cleoresult}
\eeq

Combining the errors in quadrature one has ${\rm BR}(b \rightarrow s\gamma)
=(2.32\pm 0.66)\times 10^{-4}$. The Standard Model prediction for this
branching ratio has an $O(30\%)$ uncertainty~\cite{buras}
mainly from the currently
unknown next-to-leading (NLO) order QCD corrections. Recent Standard Model
evaluations give~\cite{buras} ${\rm BR}(b \rightarrow s\gamma)
=(2.9\pm 0.8)\times 10^{-4}$ at $m_t\approx 170 GeV$. The SUSY effects in
 BR($b \rightarrow s\gamma$) can be conveniently parametrized by introducing
 the parameter
$r_{SUSY}$ which we
define by the ratio~\cite{dark}

\beq
r_{SUSY} = BR(b\rightarrow s\gamma)_{SUSY}/BR(b\rightarrow
s\gamma)_{SM}~.
\label{rsusyeqn}
\eeq
\noindent
Several uncertainties that are present in the individual branching ratios
cancel out in
the ratio $r_{SUSY}$.  However, we point out that the
NLO corrections would
in general be different for the SUSY case than for the SM case due to the
presence of different SUSY thresholds~\cite{anlauf}. In the present analysis
 we limit ourselves to the leading order evaluation.
In a similar fashion we can define

\begin{equation}
r_{exp} = BR(b\rightarrow s\gamma)_{exp}/BR(b\rightarrow
s\gamma)_{SM}~.
\label{rexpteqn}
\end{equation}
\noindent
Using the experimental result of Eq.~(\ref{cleoresult}) and the
Standard Model values given above one finds that $r_{exp}$ lies
in the range
\noindent
\begin{equation}
r_{exp}=0.46-2.2
\label{rexpvalue}
\end{equation}
\noindent
Now normally in SUSY theory one can get rather large deviations from the SM
results so that
 $r_{SUSY}$ can lie in a rather large range, i.e.,  $\approx (0,10)$.
Thus the constraint $r_{SUSY}=r_{exp}$ is an important constraint
on the theory.
  This constraint then excludes a part of the parameter space~\cite{wu,dark}
and reduces the
magnitude of the $\maxmodamu$ for a given $m_{\tilde g}$.

 The cosmological constraint on the neutralino relic density~\cite{relic}
 also plays
a very significant role in limiting the parameter space of the model.
  Theoretically the quantity of interest is $\Omega_{\tilde Z_1} h^2$ where
$\Omega_{\tilde Z_1}={{\rho_{\tilde Z_1}} \over \rho_c}$.
Here $\rho_{\tilde Z_1} $ is the
neutralino mass density, $\rho_c$ is the critical
mass density to close the universe
and $h = H/(100 km/s Mpc)$ where H is the Hubble constant.
Astronomical observations indicate $ h \cong 0.4-0.8 $
which results in a spread of value for $\Omega_{\tilde Z_1} h^2 $ .
For our analysis we use a mixture of cold and hot dark
matter (CHDM) in the ratio
$\Omega_{CDM}:\Omega_{HDM}$=2:1 consistent with the COBE data. Assuming
total $\Omega =1$ as is implied by the inflationary scenario and using for
the baryonic matter $\Omega_B$=0.1 we get~\cite{dark,relic}
\beq
0.1 \leq \Omega_{\tilde Z_1} h^2 \leq 0.4
\label{omegaeqn}
\eeq

The combined effects of the $b\rightarrow s\gamma$ constraint
and relic constraint put severe limits on the parameter
space~\cite{dark,borzumati,diehl}. Their effect on $a_{\mu}^{SUSY}$ is
 shown in Figs.~(2c) and (2d) which are similar to  Figs.~(2a) and (2b) except
for the inclusion of the combined effects of  $b\rightarrow s\gamma$ constraint
and relic constraint.
Comparison of Figs.~(2a) and (2c) and similarly of Figs.~(2b) and (2d) show
that typically $\maxmodamu$ for large $\tanbeta$ ( i.e., $\tanbeta \gsim 2$)
 is reduced by about a factor of $2 \over 3$ for
gluino masses below the dip under the
combined effects of  $b\rightarrow s\gamma$ and dark matter constraints.
However, as is obvious, the most striking effect arises due to the appearance
of the dip itself. The existence of such a dip was
first noted in Ref.~\cite{dark} and
 is due to the relic density constraint.
It is caused by the rapid annihilation of neutralinos
near the Z pole which reduces the relic density below the lower
limit in Eq.~(\ref{omegaeqn}) and hence part of the
parameter space gets eliminated due
to this constraint. In order to get the correct position and depth for
this dip one must use the accurate method ~\cite{relic,griest} for
calculation of the relic density.
The analysis of Figs.~(2c) and (2d) show that it would
be difficult to discern SUSY effects in the $\gmin2$ experiment for
gluino masses around the dip or correspondingly for neutralino masses of
$m_{\tilde Z }\sim M_Z/2 $. ( There is a similar dip at
 $m_{\tilde Z }\sim m_h/2 $ which does not appear in the graphs because
the remaining parameters have been allowed to range over the full space
 , but would be manifest once
the Higgs mass is fixed.)

Interestingly, even the current limits on $\gmin2$ including the
present experimental and theoretical errors place some
constraint on the parameter space of supergravity grand unification.
 Ascribing any new physics
to supersymmetry by using $\amu^{SUSY}+\amu^{SM}=\amu^{exp}$
one may constrain the parameter space of the model in the
$(m_0-m_{\tilde g})$ plane for different $\tanbeta$ with
a consideration of all possible $A_t$.
We have combined the uncertainty of theoretical estimate
  and the current experimental uncertainties in quadrature
to find that

\beq
-11.7 \times 10^{-9} <\amu^{SUSY}< 22.5 \times 10^{-9}
\label{limiteqn}
\eeq

Fig.~(2e) exhibits for the case $\mu < 0$ the excluded regions in
the $(m_0-m_{\tilde g})$ plane for different $\tanbeta$ values where
the excluded domains lie below the curves. We note that the excluded
domains depend strongly on the value of $\tanbeta$ and constraints
become more severe as $\tanbeta$ increases. For $\tanbeta \lsim 10$
the constraints on $(m_0,m_{\tilde g})$ are very modest. A
similar analysis
holds for $\mu>0$ as shown in Fig.~(2f), except that here the constraints
on $(m_0,m_{\tilde g})$ are generally less severe.

  We analyze now the effect of the implications of the predicted
experimental accuracy  of $\amu$
$(\sim 4 \times 10^{-10})$ to be attained in the Brookhaven experiment for
supergravity grand unification. Of course,
 the present theoretical uncertainty in the hadronic contributions to
$\amu$, mostly arising from the hadronic vacuum polarization effects as
discussed earlier,
limits the usefulness of such a precise measurement of $\amu$. The hadronic
uncertainty arises from the uncertainty in the computation of~\cite{hughes}

\beq
 R(s)={{\sigma_{tot}(e^{+}e^{-}\rightarrow hadrons)} \over
 {\sigma(e^{+}e^{-}\rightarrow \mu^+ \mu^-)}}
\label{rseqn}
\eeq
for the low energy tail of $(e^{+}e^{-}\rightarrow hadrons)$ cross-section.
Ongoing measurements in VEPP-2M together with future experiments in
DA$\Phi$NE, BEPC etc. are expected to reduce this hadronic uncertainty  to a
significant extent, perhaps by a factor of 4 or more, enhancing
correspondingly  the usefulness of the precision measurement of $a_{\mu}$.

We present here two analyses.
   For the first analysis we used the more optimistic estimate
of  Ref.~\cite{martinov} where the authors made an improved
evaluation of R(s) and $\amu^{had}$ through the use of global analytical
models of pion and kaon form factors in addition to the
use of a better experimental input of the three-pion  $e^+e^-$ annihilation
data in comparison to previous determinations~\cite{km}.
This corresponds to Case~A of Table~1 and gives (setting $R_b=0$) the result
\beq
\amu^{SM}=11659183(7.6) \times 10^{-10}
\label{smnew}
\eeq
For the second analysis we shall make a comparative study over an assumed
range of errors which includes analyses of both cases A and B of Table~1 as
well as the possibility of even more constrained errors.

  In order to analyze the effect of the predicted accuracy level
of the Brookhaven experiment on the models of our discussion
we have assumed that the experiment will not discard the Standard
Model result within its $2\sigma$ uncertainty limit. As in the analysis
of the current experiment above we ascribe any new physics
to supersymmetry by using $\amu^{SUSY}+\amu^{SM}=\amu^{exp}$
and  constrain the parameter space of our model in the
$(m_0-m_{\tilde g})$ plane for different $\tanbeta$ with
a consideration of all possible $A_t$.
Following the same procedure as before, we combine the uncertainty of
theoretical estimate
  and the expected experimental accuracy
level of $4 \times 10^{-10}$ in quadrature.
In our first analysis we use Eq.~(\ref{smnew}) as
the theoretical input and assume that the predicted accuracy
in the experimental determination of $\amu$ will be achieved, we then
determine the constraints on the SUSY particle spectrum if  $\amu^{SUSY}$ lies
within the $2\sigma$ of the combined theoretical and experimental error.
  In Figs.~(3a) to (3f) we exhibit the regions of
 $(m_{\tilde W_1}-m_{\tilde \nu_\mu})$
plane which will be excluded (dark shaded region) if $\amu^{SUSY}$
lies within the  $2\sigma$ limit.We also exhibit the regions which will
be partially excluded (light shaded  area) because a significant part of
the parameter space is eliminated by the constraint, and the allowed region
 (dotted area). In addition, there is a region (white space) which is
inaccessible due to the existence of a lower limit
of sneutrino mass $(m_{\tilde \nu_\mu})$~\cite{dark}.

Next we give a comparative analysis of the constraints for cases~A and B of
Table~1 and also for a case~C where the error is reduced
by factor of 4 over case~B ( see in this context the analysis of
Ref.~\cite{Dubnickova} which gives a new evaluation of hadronic
contributions of $699(4.5) \times 10^{-10}$ ).
In Figs.~(4a) and (4b), we give a composite display of the excluded regions
in the $(m_0-m_{\tilde g})$ plane for the three cases.
We observe that the forbidden region increases
in proportion to the decrease in the combined error of theory and experiment.
Further, the excluded region increases  with increasing value of $\tanbeta$.
Thus if the combined error decreases by a factor of four (case~C) and no effect
beyond $2\sigma$ is seen, then the ${\rm g-2}$ experiment will exclude most of
the  region in $(m_0-m_{\tilde g})$ plane for large $\tanbeta$ as can be seen
from Figs.~(4e) and (4f). In fact even with the presently large uncertainty
in the theoretical values of   $\amu^{SM}$ one will be able to exclude
a significant part of the $(m_0-m_{\tilde g})$ parameter space if the expected
sensitivity of the measurement of $\amu$ is reached for large $\tanbeta$
values (see Figs.~(4e) and (4f)).
Of course a significant reduction of the
 uncertainty in $\amu^{SM}$ which one expects to be possible
in the near future will more stringently constrain the parameter space
when combined with the expected sensitivity
 of the Brookhaven experiment.
 We have also carried out a similar analysis for the $a_{\tau}^{SUSY}$ for
the tau lepton. This gives $max~|a_\tau^{SUSY}| \sim 1.0 \times 10^{-5}$ for
 $\tanbeta =20$. Even for this large value of $\tanbeta$ the predicted
value of $a_\tau^{SUSY}$ is too small to observe with present
experimental accuracy~\cite{taupapers}.

\section{Conclusion}

In this paper we have presented an analysis of $\gmin2^{SUSY}$
within the framework of supergravity grand unification under the
constraint of radiative breaking of the electro-weak symmetry, and
the constraint of $b\rightarrow s\gamma$ and dark
matter. One finds that over most of the parameter space the chiral
interference light chargino part of  $a_{\mu}^{SUSY}$ dominates and
 imparts a signature to  $a_{\mu}^{SUSY}$.Thus as a rule one finds
$a_{\mu}^{SUSY}>0$ for $\mu>0$ and
$a_{\mu}^{SUSY}<0$ for $\mu<0$  with very few exceptions when
tan$\beta\sim 1$. At the quantitative level it is shown that
 the expected experimental sensitivity
of $\gmin2$ measurements combined with the expected reduction of error
in $(g-2)_\mu^{SM}$ ( by $\approx O(1/4)$)  will exclude
the $m_0, m_{\tilde g}$ parameters in the domain $m_0\lsim 700 $ GeV,
$m_{\tilde g}\lsim 1 $ TeV for large $\tanbeta (\gsim 20)$
and stringently constrain the parameter space for lower $\tanbeta$.
With the same assumptions one will be able to probe
 the domain $m_0\lsim 400$ GeV,
$m_{\tilde g}\lsim 700$ GeV for  $\tanbeta \gsim 10$.  The constraint
becomes less severe for smaller values of $\tanbeta$. However, even for
$\tanbeta=5$, the SUSY reach of the new experiment will be very
substantial (see Figs.~(4a) and (4b)). Thus one finds that the Brookhaven
experiment coupled with the corroborating experiments and analyses
designed to reduce the hadronic error will complement SUSY searches at
colliders and  provide an important probe
of the parameter space of supergravity grand unification especially
for large $\tanbeta$.

 We wish to acknowledge useful discussions with R. Arnowitt, W.~Marciano and
W.~Worstell. This research was supported in part by NSF grant number
PHY--19306906 and PHY94-07194.

\section{ Appendix}

  The chargino masses $m_{\tilde W_i}=|{\lambda}_i|$, i=1,2
 where $\lambda _{1,2}$ are the eigenvalues of the chargino mass matrix
and are given by
\beq
{\lambda_{1,2}}={\frac{1}{2}}\left(
{\left[4{\nu}^2_2+{({\mu}-{\tilde m}_2)}^2 \right]}^{\frac{1}{2}} \mp
{\left[4{\nu}^2_1+{({\mu}+{\tilde m}_2)}^2 \right]}^{\frac{1}{2}} \right)
\label{lambdaequation}
\eeq
where
\beq
{\nu}_{1,2}={{M_W} \over {\sqrt2}}{{\left(\sin {\beta} \mp \cos
{\beta}\right)}}
\eeq
and
${\tilde m}_2$ is obtained from the relation
${\tilde m}_a={{\alpha_a(M_Z)} \over {\alpha_3(M_Z)}}m_{\tilde g}$
where $\alpha_a(M_Z)$, $a=1,2,3$ are the SU(3), SU(2) and U(1) gauge
coupling constants at the Z boson mass.

The neutralino masses $m_{\tilde Z(k)}=|\lambda _k|$ where
$\lambda _k$ are the eigenvalues of the neutralino mass matrix which in
the ($\tilde{W}_{3}, \tilde{B}, \tilde{H}_{1},
\tilde{H}_{2}$) basis reads~\cite{chams}

\begin{equation}
M_{\tilde{Z}}=\pmatrix{\offinterlineskip
{\tilde{m}_{2}}&o&\vrule\strut&a&b\cr
o&{\tilde{m}_{1}}&\vrule\strut&c&d\cr
\noalign{\hrule}
a&c&\vrule\strut&o&{-\mu}\cr
b&d&\vrule\strut&{-\mu}&o\cr}
\end{equation}

\noindent
where $ a=M_{Z}
\cos\theta_{W}\cos\beta,~ b=-M_{Z}\cos\theta_{W}\sin\beta,~
c=-M_{Z}\sin\theta_{W}\cos\beta$ and $d=M_{Z}\sin\theta_{W}\sin\beta$ while the
quantities $\theta_k$ that appear in Eq.~(\ref{brbleqn})
are defined by $\theta_k=0~(1)$ for $\lambda_k>0~(<0)$.
 The quantities $O_{ij}$ are the elements of the orthogonal matrix which
diagonalizes the neutralino mass matrix.

Smuon masses are given by
\beq
m^2_{\tilde \mu_{1,2}}={\frac{1}{2}}
({\left({{\tilde m}^2}_L+{{\tilde m}^2}_R \right)
 \mp \sqrt{ {\left({{\tilde m}^2}_L-{{\tilde m}^2}_R \right)}^2
        + 4m_\mu^2 {\left(A_t +\mu \cot\beta \right)}^2}})
\eeq
Here $A_t$ is scaled with $m_0$, and the L and R parts are given by
\begin{eqnarray}
\tilde m^2_L & = & m_0^2 +m_\mu^2 +\tilde \alpha_G
\left[{\frac{3}{2}}f_2 +{\frac{3}{10}}f_1\right]m_{\frac{1}{2}}^2
+\left(-{\frac{1}{2}}+\sin^2{\theta_W}\right)M_Z^2\cos2\beta  \nonumber \\
\tilde m^2_R & = & m_0^2 +m_\mu^2 +\tilde \alpha_G
\left({\frac{6}{5}}f_1\right)m_{\frac{1}{2}}^2
-\sin^2{\theta_W}M_Z^2\cos2\beta
\end{eqnarray}
where\hspace{2mm} $m_{\frac{1}{2}}={{\alpha_G} \over {\alpha_3(M_Z)}}
m_{\tilde g}$,
$ \tilde \alpha_G = \alpha_G/4\pi$, $f_k(t) = t(2+\beta_kt)/(1 +
\beta_kt)^2$
with $\beta_k ={\tilde \alpha_G}(33/5, 1, -3)$ and
$t=2{\rm ln}({{M_G} \over Q})$ at $Q=M_Z$.

  The mixing angle which describes the left-right mixing for the smuons is
determined by the relations
\beq
{\sin2\delta}={{2m_\mu \left({A_t}+{\mu}\cot\beta \right)} \over
{\sqrt{\left(\tilde m^2
_L-\tilde m^2_R \right)^2 + 4m_\mu^2
\left(A_t+ {\mu}\cot\beta\right)^2}}}
\label{deleqn}
\eeq
and
\beq
{\tan2\delta}={{2m_\mu\left({A_t}+{\mu}\cot\beta \right)} \over
{\left(\tilde m^2_L- \tilde m^2_R \right)}}
\label{tandel}
\eeq

A similar result analysis holds for sneutrino masses and one has
\beq
m^2_{\tilde \nu_\mu}=m_0^2 +\tilde \alpha_G
\left[{\frac{3}{2}}f_2 +{\frac{3}{10}}f_1\right]{m_{\frac{1}{2}}}^2
+{\frac{1}{2}}M_Z^2\cos2\beta
\label{sneutrinoeqn}
\eeq

\newpage
\begin{center}
{\large\bf Figure Captions}
\end{center}
\vspace{0.5cm}
\noindent
{\bf Fig.~1:}\\
 Fig.~(1a) Chargino-sneutrino one-loop exchange diagram which
contributes to
$a_\mu^{SUSY}$.\\
Fig.~(1b) Neutralino-smuons one-loop exchange diagram which
contributes to $a_\mu^{SUSY}$.\\

\noindent
{\bf Fig.~2:}\\
Fig.~(2a) The upper limit of
$|a_\mu^{SUSY}|$ {\it vs.} $m_{\tilde g}$ for the case $\mu< 0$
in the Minimal Supergravity model for different values of $\tanbeta$
in the region $\tanbeta$ $\leq 30$, when one allows the remaining
parameters $A_t$ and $m_0$ to vary over the parameter space subject to
the naturalness criterion of $m_0 \leq$~1~TeV.\\
The dashed horizontal line is the current $2\sigma$ limit as given by
Eq.~(\ref{limiteqn})\\
Fig.~(2b): Same as Fig.~(2a) except that $\mu> 0$.\\
Fig.~(2c): Same as Fig.~(2a) except that the constraint from
$b\rightarrow s\gamma$ and dark matter as discussed in the text
are included.\\
Fig.~(2d): Same as Fig.~(2c) except that $\mu> 0$.\\
Fig.~(2e): Excluded regions (regions enclosed by the curves towards the
origin)
in the  $(m_0-m_{\tilde g})$
plane under the current theoretical and experimental limits of $a_\mu$
  for various
values of $\tanbeta$, for the case $\mu< 0$ .\\
Fig.~(2f): Same as Fig.~(2e) except that $\mu> 0$.\\

\noindent
{\bf Fig.~3:}\\
Fig.~(3a) Display in the $(m_{\tilde W_1}-m_{\tilde \nu_\mu})$
plane of the allowed (shown in dots), disallowed
(shown in squares) and partially allowed (shown in cross)
regions corresponding to the 2$\sigma$ limit of Case~A of the
theoretical evaluation of $a_\mu$ and the predicted level of
accuracy of the Brookhaven experiment, for the case tan$\beta=5$ and
$\mu< 0$.
The white area between the excluded region and $m_{\tilde W_1}$ axis
remains inaccessible due to the existence of a lower limit of sneutrino mass
$m_{\tilde \nu_\mu}$~\cite{dark}.\\
Fig.~(3b): Same as Fig.~(3a) except that $\mu>0$.\\
Fig.~(3c): Same as Fig.~(3a) except that tan$\beta=10$.\\
Fig.~(3d): Same as Fig.~(3c) except that $\mu>0$.\\
Fig.~(3e): Same as Fig.~(3c) except that tan$\beta=30$.\\
Fig.~(3f): Same as Fig.~(3e) except that $\mu> 0$.\\

\noindent
{\bf Fig.~4:}\\
Fig.~(4a): Excluded regions (regions enclosed by the curves towards the
origin)
 for $\tanbeta=5$ and $\mu<0$ in the 2$\sigma$ limit, after combining
in quadrature the predicted Brookhaven
experimental uncertainty
and different levels of uncertainty of theoretical estimates
corresponding to Cases~A, B, and C as discussed in the text.\\
Fig.~(4b): Same as Fig.~(4a) except that $\mu> 0$.\\
Fig.~(4c): Same as Fig.~(4a) except that $\tanbeta$=10.\\
Fig.~(4d): Same as Fig.~(4c) except that $\mu> 0$.\\
Fig.~(4e): Same as Fig.~(4a) except that $\tanbeta$=30.\\
Fig.~(4f): Same as Fig.~(4e) except that $\mu>0$.\\

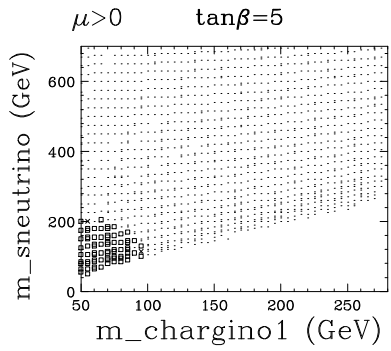
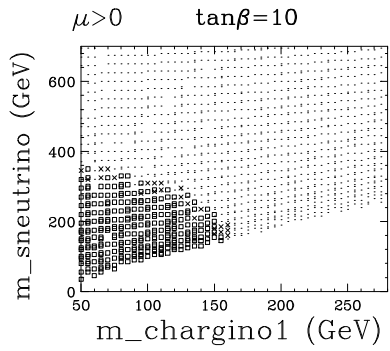
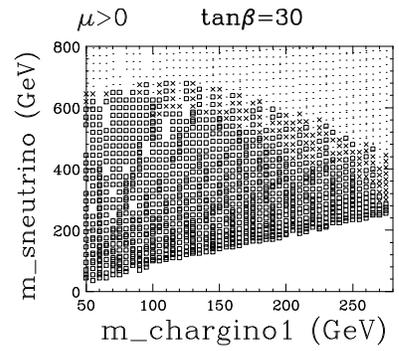
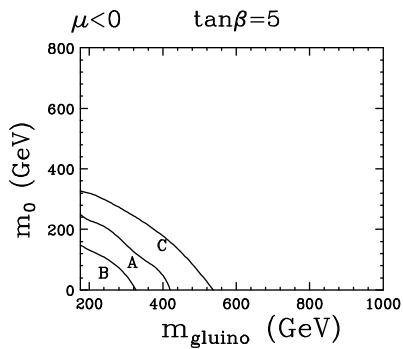
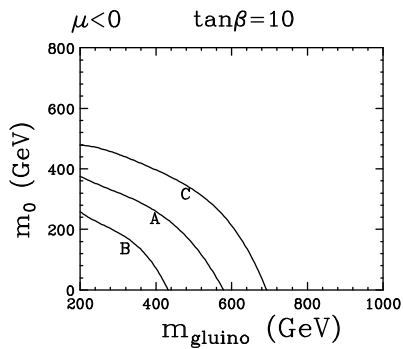
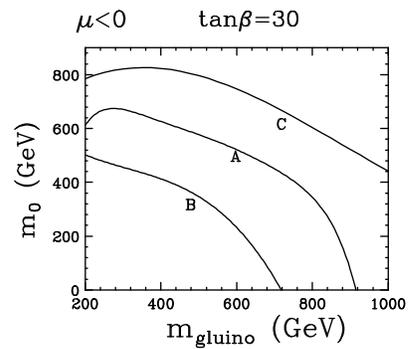
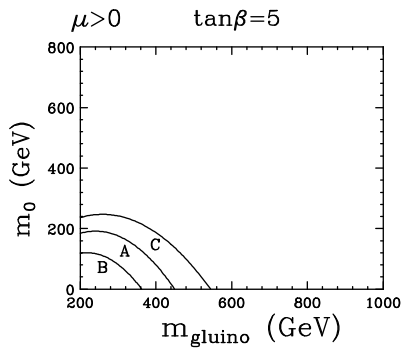
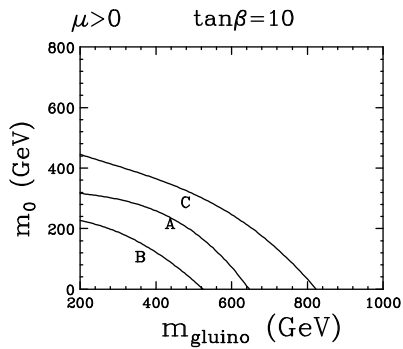
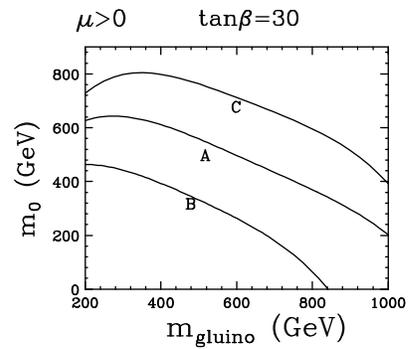
\end{document}